\documentstyle[11pt,aaspp4]{article}

\def\et{et al.}

\def\ha{H$\alpha$}

\def\rhalf{R$_{0.5}$}

\def\solar{\ifmmode_{\mathord\odot}\;\else$_{\mathord\odot}\;$\fi}

\def\HII{H$\,${\sc ii}}
\def\HI{H$\,${\sc i}}

\begin{document}

\title{The Stellar Population and Star Clusters in the Unusual Local
Group Galaxy IC 10\footnote{\rm Based
on observations with
the NASA/ESA  Hubble Space Telescope, obtained at the Space Telescope
Science Institute, which is  operated  by the Association of Universities
for Research in Astronomy, Inc. 
under NASA contract NAS5-26555.}
}

\author{Deidre A.\ Hunter}
\affil{Lowell Observatory, 1400 West Mars Hill Road, Flagstaff, Arizona 86001
USA;
\\dah@lowell.edu}

\begin{abstract}

We present analysis of {\it Hubble Space Telescope} U, V, I, and \ha\ 
images of the peculiar Local Group 
irregular galaxy IC 10. The images are used to determine the nature
of the stellar products in a portion of the recent 
starburst in this galaxy.  
We identified 13 stellar associations and clusters, two of which
are probably old ($\geq$350 Myr) 
and the rest of which are young (4--30 Myr) and presumably formed
in the starburst.
We found the following:
1) The slope of the 
stellar initial mass function (IMF) 
for 6.3--18 M\solar\ 
stars formed in the starburst
lies between two limiting cases: a value of 
$-1.9\pm0.4$ 
under the assumption of coevality
over the past 13 Myr and of $-0.9\pm0.3$ under the assumption of 
constant star formation over the past 40 Myr for $Z=0.004$
($-2.1\pm0.4$ and $-1.0\pm0.04$, respectively, for $Z=0.008$).
Thus, most likely, the IMF of the intermediate mass stars is not
very unusual.
The slope of the IMF for the 
underlying galaxy population under the assumption of constant
star formation is $-2.6\pm0.3$
for 4.8--18 M\solar\ stars
assuming $Z=0.004$ ($-2.3\pm0.3$ for $Z=0.008$), and is unusually steep.
2) The lower stellar mass limit in the starburst is $\leq6.3$ M\solar.
This constraint is less than some predictions of what lower stellar mass limits
might be in starbursts, but higher than others.
3) There are 
two modest-sized \ha\ shells ($\sim$50 pc diameter) that could
easily have been produced
in the past few Myr by the clusters they encircle.
4) 
The dominant mode of star formation in the starburst has been that 
of a scaled-up
OB association. 
This mode, with a few compact clusters sprinkled in, is similar
to the star formation that took place in Constellation III in
the LMC, as well as that in the Blue Compact Dwarfs IZw18 and VIIZw403.
The starburst in this part of IC 10 has not produced a super star cluster.
We also compare the high WC/WN ratio 
to evolutionary models and discuss possible explanations.
The high ratio can be reproduced if there were
small, well-synchronized ($\Delta\tau\leq1$ Myr), 
but widely scattered, pockets of secondary star formation
3--4 Myr ago.

\end{abstract}

\keywords{galaxies: irregular --- galaxies: star formation
--- galaxies: individual: IC 10 ---
galaxies: star clusters --- Local Group --- galaxies: starburst
--- galaxies: stellar content}

\section{Introduction}

IC 10 is a small irregular galaxy in the Local Group. With an
M$_B$ of $-$16.5, IC 10 has an integrated
luminosity that is comparble to that of the SMC
(although IC 10 is 0.2 magnitude redder in (B$-$V)$_0$ and the same in
(U$-$B)$_0$; de Vaucouleurs \et\ 1991). 
What is unusual about IC 10 is that it is undergoing a 
wide-spread burst of star formation, according to a study of the
massive stars by Massey \& Armandroff (1995).
Massey \& Johnson (1998) found that the density of evolved massive
stars in the Wolf-Rayet phase is several times higher than that of the LMC,
but the LMC is 4 times more luminous than IC 10.
The integrated star formation rate for IC 10 inferred from the \ha\ luminosity is
0.03 M\solar\ yr$^{-1}$ kpc$^{-2}$ and is high compared to most other
irregular galaxies (Hunter 1997).
(Note: The rate derived by Borissova \et\ (2000) from \ha\ and
Br$\gamma$ imges, normalized to the area of the galaxy is 8 times higher
than this value,
primarily due to higher extinction corrections used).

IC 10 is also known to be unusual in having \HI\ gas that extends about 7 times
the optical dimensions of the galaxy (Huchtmeier 1979).
Furthermore, that extended gas contains a large \HI\ cloud
(Wilcots \& Miller 1998), which it has been
suggested is falling into the galaxy causing the current heightened
star formation activity (Sait\=o \et\ 1992).
At a distance of $\sim$0.5---1 Mpc (Massey \& Armandroff 1995;
Saha \et\ 1996; Wilson \et\ 1996; Sakai, Madore, \& Freedman 1999;
Borissova \et\ 2000),
IC 10 is the nearest
example of the starburst phenomenum. As such,
it offers a unique opportunity to test ideas about the star
formation process in a starburst environment by examining the products
of the starburst event.
 
In studies of a few
starburst galaxies, some people have come to the conclusion that lower
mass stars must not have formed in the recent round of intense star
formation in those systems (Scalo 1987, 1990). In M 82, for example,
Rieke \et\ (1993) and Doane \& Mathews (1993) argue for a deficit of
stars of mass less than a few M\solar. Another interacting system
(Mrk 171,IC 694) and a Blue Compact Dwarf (Tol 65) may have
formed only O or early B stars (Augarde \& Lequeux 1985, Olofsson 1989),
and a third interacting system is believed to have a lower mass limit,
$M_l$, of 3--6 M\solar (Wright \et\ 1988).
 
Unfortunately, because these galaxies are at large distances, individual
stars cannot be resolved and the arguments for high $M_l$'s have
necessarily been indirect and, therefore, highly uncertain. In fact
Scalo (1990) reviews the evidence for unusually high $M_l$'s and
concludes that many starburst galaxies do not seem to require a
top-heavy stellar initial mass function (IMF).
Furthermore, in R136, the super star cluster in the LMC, where star
formation has been {\it locally} intense, the stellar IMF 
is normal down to at least 3 M\solar\ (Hunter \et\ 1995,
Sirianni \et\ 2000)
although massive star clusters near the Galactic center {\it do} 
appear to have an unusually shallow IMF for stars with masses greater
than 10 M\solar\ (Figer \et\ 1999).
The case for a top-heavy IMF in {\it some} systems is hard
to dismiss.

Theoretical arguments also bolster the expectation that $M_l$ could be
unusually high in environments of intense star formation, even in
regions of galaxies that are only locally intense. For example, Silk
(1977) and Larson (1985) have suggested that massive stars form in
hotter, more turbulent environments than do lower mass stars. Therefore,
G\"usten \& Mezger (1983) concluded that $M_l>$2--3 M\solar in
spiral arms, and Silk (1986) has suggested that $M_l$ could be as high as 10
M\solar\ in giant \HII\ regions.
 
People have long been suspicious that not only
is $M_l$ unusual in a starburst, but that the proportions of stars
that are formed are unusual as well.
In IC 10, Massey \& Armandroff (1995) 
found an unusually large ratio of the evolved massive Wolf-Rayet
WC-type to WN-type stars
for this galaxy, a ratio that
is 20 times too high for its metallicity (Massey 1999).
They suggested that IC 10 may 
have an IMF that is skewed towards the highest
mass stars, which they blamed on the very vigorous star formation
occurring there now. 

Besides the proportions and mass limits of the stars,
the nature of the star clusters formed in a starburst is also a clue
to the conditions for star formation that were endemic during the burst.
For example, some starbursts in small galaxies
have produced super star clusters---
young, compact, luminous clusters, many of which may be
as massive as globular clusters in the Milky Way.
These include the irregulars NGC 1569, NGC 1705, and NGC 4449
(Meurer \et\ 1992;
O'Connell, Gallagher, \& Hunter 1994;  Ho \& Filippenko 1996;
Gelatt, Hunter, \& Gallagher 2001),
as well as the giant star-forming event 30 Doradus in
the LMC (Hunter \et\ 1995) and several embedded sources inferred to be
super star clusters (NGC 2363 in NGC 2366---Drissen \et\ 2000;
He2-10---Conti \& Vacca 1994, Kobulnicky \& Johnson 1999;
NGC 4214---Leitherer \et\ 1996;
NGC 5253---Turner, Ho, \& Beck 1998).
Other starbursts, however, have not produced super star clusters
(IZw18---Hunter \& Thronson 1995; VIIZw403---Lynds \et\ 1998).
Thus, we would like to know what kinds of star clusters have
been produced in the recent starburst in IC 10.

Because of its relatively close proximity, therefore, we chose to use 
IC 10
as a test for star formation
processes in a starburst galaxy. Towards this end, we obtained 
{\it Hubble Space Telescope}
({\it HST}) images of IC 10 through three broad-band filters that
simulate U, V, and I and a narrow-band \ha\ filter.
These data allow us to examine the products of the star formation
process,
particulary the intermediate mass stars and star clusters.
We are addressing the questions: 
1) What is the stellar IMF of the intermediate mass stars produced
in the starburst?
2) What is the lower stellar
mass limit of stars produced in the recent starburst?
3) What kinds of star clusters have been produced in the starburst?
4) How does the nebular emission relate to the stellar products?
5) What has been the mode of star formation in the starburst?
In a subsequent paper we will examine constraints on the galactic star 
formation history
from the color-magnitude diagram of the stellar population.
 
\section{Observations and Data Reduction}

IC 10 was imaged with the Wide Field and
Planetary Camera 2 (WFPC2) on {\it HST} in two sessions.
Observations through filters F336W and F814W were obtained on 
1997 June 8, and observations through filters F555W and F656N
were obtained at the same orientation on 1999 June 9 and 10.
The WFPC2 camera consists of a PC CCD with a resolution of
0.0456\arcsec\ per pixel, which is 0.22 pc at the galaxy, and
three WF CCDs with a resolution of 0.0996\arcsec\ per pixel,
which is 0.48 pc at the galaxy.
We obtained three 800 s exposures through each of filters F336W and F656N 
and 10 1400 s exposures through each of filters F555W and F814W.
The multiple exposures were combined to
remove cosmic rays but conserve flux.
 
Basic data reduction steps were done by the Space Telescope Science
Institute ``pipeline'' processing system.
We produced a nebular emission image by combining the
F555W and F814W images, shifting, scaling,
and subtracting from the F656N image to remove the stellar
continuum.
We also produced nebular-free F555W and F814W images by scaling and
subtracting the nebular emission image.
The field of view of the {\it HST} images is shown superposed on
a ground-based image of IC 10 in Figure \ref{figfov}.
 
We measured the brightnesses of stars in the galaxy
using the crowded star photometry package HSTphot, designed by 
Dolphin (2000a) after
DAOPHOT (Stetson 1987) but optimized for WFPC2 images.
Stars were eliminated if they had 
high $\chi$ and deviant sharpness parameters,
measures of the shape of the object relative to the input 
point-spread-function.
The photometry was corrected for the Charge Transfer Efficiency problem
that causes some signal to be lost when charge is transferred down the
chip during readout. This problem affects objects at higher row
numbers more than those at lower row numbers and is a function of
time elapsed since WFPC2 was installed. We used the formulism 
for correcting
for this problem and the calibration parameters given by Dolphin (2000b).
We also converted the F336W, F555W, and F814W
photometry to the Johnson U and V and Kron-Cousins I systems using the
conversions of Dolphin (2000b), which are minor adjustments to the original
calibration by Holtzman \et\ (1995). The instrumental magnitudes
were corrected for reddening and the red leak in F336W, as discussed
below, before
converting to U, V, and I, as discussed by Holtzman \et\
We required that stars be successfully measured in both F555W and
F814W to be retained, and the final photometry list contains
44241 stars.

We also used ground-based V and R-band images of IC 10 given to us 
by P.\ Massey.
The images were taken in 1992 at Kitt Peak National Observatory with 
the 4 m telescope and
a Tektronix 2048$\times$2048 CCD.
There were two exposures of 500 s each. The field of view was 14.3\arcmin.

\section{Data Analysis Issues}

\subsection{Reddening} \label{secred}

IC 10 lies in the plane of the Milky Way and so is highly reddened from
foreground extinction.
In addition there is likely to be some
internal reddening within IC 10.
Estimates of the total reddening E(B$-$V)$_t$ 
to and within IC 10 have varied greatly.
De Vaucouleurs \& Ables (1965) used integrated U, B, and V colors of the galaxy to
estimate E(B$-$V)$_t$ at 0.87.
Lequeux \et\ (1979) measured the Balmer decrement from emission-line
spectroscopy to derive a reddening of 0.47.
Yang \& Skillman (1993) used the ratio of \ha\ to radio continuum
fluxes in two \HII\ regions to find E(B$-$V)$_t$ of 1.7--2.
Massey \& Armandroff (1995) estimate the reddening at 0.75--0.80 from
Wolf-Rayet (WR) stars and from a comparison of the blue main sequence to that
of the LMC.
Saha \et\ (1996) determined an
E(B$-$V)$_t$ of 0.94 from the blue limit for unevolved supergiants.
Wilson \et\ (1996) determined the reddening at 0.8 from near-infrared
observations of Cepheids.
Sakai \et\ (1999) also analyzed near-infrared observations of a few 
Cepheids along with V,I photometry of others to conclude that 
E(B$-$V)$_t=1.16\pm0.08$.
Borissova \et\ (2000) compare JHK photometry of red supergiants in IC 10
to those in IC 1613 and derive 
E(B$-$V)$_t=1.05\pm0.10$, with higher extinction to \HII\ regions.
The compilation of Schlegel, Finkbeiner, \& Davis (1998) used to map
dust in the Milky Way concludes that the {\it foreground} reddening
at the position of IC 10 is 1.6 in E(B$-$V). Thus, estimates of E(B$-$V)$_t$
for the stars in IC 10 range from 0.5 to somewhat greater than 1.6 magnitudes.

We began by adopting an E(B$-$V)$_t$ of 0.77 from Massey \& Armandroff
(1995), and dereddened the stellar photometry using this.
In the U, V, and I color-magnitude diagrams (CMDs)
of the burst region and of individual star clusters (presented
in \S\ref{seccmd}), we
found that isochrones fall along the blue edge of the blue plume as
expected. The agreement between the model main sequence and the
observed photometry is quite good, suggesting that E(B$-$V)'s much
higher than 0.77 would not be reasonable.
In particular, an E(B$-$V)$_t$ of 1.6 magnitude would require that the blue
plume and blue luminous cluster stars have a (V$-$I)$_0\sim -1.3$
and (U$-$V)$_0\sim -2.9$, which are unphysical.
An E(B$-$V)$_t\sim 1$, as found by several authors, would move the
middle of the blue plume 0.1 magnitude 
and the blue edge 0.3 magnitude blueward of the isochrones
in (V$-$I)$_0$ and similarly in (U$-$V)$_0$.
The stars in the resolved clusters would also fall blueward of the isochrones
in the CMDs.
Thus, while there is certainly some uncertainty in the reddening
adopted here, an overall reddening of the stars in IC 10 as high as 1.0
magnitude in E(B$-$V)$_t$ is inconsistent with our photometry. However,
an additional E(B$-$V) of order 0.05 magnitude would bring the isochrones
to the middle of the blue plume and the red supergiants more in line with
the ends of isochrones (but see discusssion in \S\ref{seccmd}),
and so a somewhat higher E(B$-$V)$_t$ cannot be ruled out.

However, the main sequence does have breadth to it, of order 0.5 magnitude
in (V$-$I)$_0$ and 0.7 magnitude in (U$-$V)$_0$. 
Some of this must be evolution of the stars, as well as the presence
of binary stars and stellar rotation,
but could there be significant {\it variable} reddening within the 
field of view of the WFPC2?
Massey \& Armandroff (1995) examined this issue by looking at the reddenings
determined from WR stars. They divided the WR stars into two groups---one
group lying at high HI column densities, and the other at lower column
densities. The higher HI column densities might imply a higher reddening
as well. They found a reddening of 0.73$\pm$0.08 for one group and
0.87$\pm$0.05 for the other group. Thus, a variation of order 0.1
in E(B$-$V) is possible. Yang \& Skillman (1993) found a difference
in E(B$-$V) of 0.3 between two \HII\ regions from radio continuum and \ha\
observations as did Borissova \et\ (2000).
However, one would expect the reddenings within nebulae
to be higher and more variable than that experienced by unembedded stars.
We conclude that some differential reddening probably is occurring.
One might expect this in particular in regions of
residual nebulosity.
However, if some stars are reddened more than our adopted
value by $<$0.1 magnitude, the error in (V$-$I)$_0$ is of order $<$0.1.
We have, therefore, adopted the single E(B$-$V)$_t$ 
value of 0.77 for all stars,
but recognize that some scatter in the CMDs could be due to differential
reddening.

Holtzman \et\ (1995) showed that the reddening correction in the 
WFPC2 filters
is a function of the spectrum of the object (see also Grebel \& Roberts 1995),
and in F336W the difference
between the extinction of an O6 star and a K5 star can be large.
Since the colors of the stars cover a large range,
we have determined a reddening
correction that depends on the observed F555W$-$F814W color of the star.
For stars with F555W$-$F814W$<$1.25, we used the corrections appropriate
for a blue star (Holtzman \et's O6 star; $A_{336}=3.935$, 
$A_{555}=2.482$, and $A_{814}=1.472$); for stars with 
F555W$-$F814W$>$2, we used the corrections appropriate for a red star
(Holtzman \et's K5 star; $A_{336}=2.932$, $A_{555}=2.338$, and
$A_{814}=1.412$); and for stars in between these two colors
we used an average of the O6 and K5 corrections. 
Our use of these three distinct color bins results in the two gaps in 
(V$-$I)$_0$ visible in the CMDs.
We followed this same proceedure for correcting the integrated
colors of star clusters as well.

\subsection{Distance} \label{secdist}

Distance estimates to IC 10 have varied as widely as have reddening
estimates.
In their study of WR stars in IC 10, Massey \& Armandroff
(1995) derived an (m$-$M)$_0$
of 24.9. Saha \et\ (1996) derived a distance modulus of 24.59$\pm$0.30
from optical observations of Cepheids.
Wilson \et\ (1996) concluded that (m$-$M)$_0$ is 24.57$\pm$0.21
from near-infrared observations of Cepheids.
Sakai \et\ (1999) use V,I and near-infrared photometry of Cepheids to
derive a distance of 660$\pm$66 kpc and the tip of the red giant
branch (TRGB) to obtain a lower
limit to the distance of 500 kpc.
Borissova \et\ (2000) used red supergiants to obtain a distance of
590$\pm$35 kpc.

With the stellar photometry corrected for reddening as described in the
previous section, we have
estimated the distance to IC 10 from the 
TRGB following the method outlined by Lee, Freedman, \& Madore (1993).
In this method m$-$M$=$I$_{TRGB}$$+$BC$_I$$-$M$_{Bol,TRGB}$,
where I$_{TRGB}$ is the apparent I-band magnitude of the TRGB determined
from a break in the I-band luminosity function, BC$_I$ is the
bolometric correction to the I-band luminosity of the TRGB,
and M$_{Bol,TRGB}$ is the bolometric I-band luminosity of the TRGB.
BC$_I$ is a function of (V$-$I)$_{TRGB}$, and M$_{Bol,TRGB}$
is a function of (V$-$I)$_{-3.5}$, which is (V$-$I) at M$_I$$=$$-$3.5.
The I$_0$ versus (V$-$I)$_0$ CMD is shown in Figure \ref{figicmd}.
Stars to the right of the dashed line were taken to be red giant branch
(RGB) and asymptotic giant branch (AGB)
stars and included in the I-band
luminosity function (see, for example, Lee 1993).
We included only stars located outside the burst region and
corrected for foreground stars and incompleteness, as described below.

The I-band luminosity function is shown in Figure \ref{figlumi}.
The TRGB is characterized by an abrupt change in star counts
in this luminosity function. Various examples for other irregulars
are shown by Lee \et\ (1993),
and this method was applied to IC 10 by Sakai \et\ (1999)
using ground-based data.
In Figure \ref{figlumi} there is no
overwhelmingly obvious break. From the appearance of the CMD we expect that the
TRGB lies somewhere between 20.5 and 21 magnitudes. 
There is a small jump at 20.8 magnitudes.
If this represents the TRGB, then (V$-$I)$_{TRGB}=1.74$,
(V$-$I)$_{-3.5}=1.8$, and (m$-$M)$_0$ is 24.95$\pm$0.2. 
The uncertainty is simply taken to be the width of our luminosity bins.
The distance modulus obtained by Sakai \et\ (1999) from the TRGB
is 24.1, corrected to an E(B$-$V) of 0.77 used here. Thus, their
distance is 0.9 magnitude or 290 kpc closer. However, they consider
their TRGB distance to be a lower limit because of reddening uncertainties.

Thus, having adopted the reddening of Massey \& Armandroff (1995),
we have recovered the distance that they determined as well.
A different choice of extinction would result in a different distance,
but we have shown in \S\ref{secred} that a choice of E(B$-$V)$_t$
that is much higher leads to unphysical stellar colors.
Therefore, here we use the combination of
E(B$-$V)$_t$$=$0.77 and (m$-$M)$_0$$=$24.9 (D$=$0.95 Mpc).

\subsection{Red Leak in F336W}

We have corrected the F336W photometry for the red leak in that
filter. 
This is particularly important for objects in IC 10 which are
highly reddened from foreground extinction. 
To determine the red leak as a function of the
observed F555W$-$F814W color, we used the simulations in STSDAS
and blackbody curves reddened by a total E(B$-$V)$_t$ of 0.77.
The red leak was taken to be any flux contribution from wavelengths
$\geq$4000 \AA,
after the definition of Holtzman \et\ (1995).
The F336W stellar photometry was corrected for the red leak
based on its observed F555W$-$F814W color. 
For a red star with an observed
F555W$-$F814W of 2,
the contribution of the red leak to the F336W magnitude is 0.19
magnitude. The correction increases rapidly
for redder observed colors, but the reddest stars were also too faint
in U to detect in F336W. All but a few stars detected in F336W have
F555W$-$F814W$\leq$2.
The large scatter in (U$-$V)$_0$ for fainter stars in V,U$-$V CMDs,
shown below,
suggests that the red leak correction has not been totally successful
for stars that are faint in the F336W image.

\subsection{Star Count Incompleteness Corrections} \label{secincom}

We estimated incompleteness factors in the F555W and F814W images
as a function of magnitude and F555W$-$F814W color by adding 35 million artificial
stars one at a time to the images. Using HSTphot, we added the artificial stars,
reran the photometry and extraction proceedures as we had originally,
and determined what stars were recovered and at what magnitude they were 
measured. Stars were added according to the original numbers and 
distributions of
stars on the F555W images.
We required that a star be recovered on both F555W and F814W images.
We used these data to determine incompleteness corrections for each
region, color bin, and magnitude bin in which we were counting stars. 
Since the incompleteness corrections are determined for each region, CCD
chip, and color and magnitude binning used in the following calculations,
we cannot present all corrections here. However, in Figure \ref{figincom}
we present a representative sample of incompleteness corrections used
below.

\subsection{Subtraction of Foreground and Underlying Stellar Populations}

Because IC 10 is located at low Galactic latitude, there is a considerable
foreground star population that must be removed at least statistically.
We have used the Bahcall-Soneira model of the Galaxy
(Bahcall \& Soneira 1980) to estimate the contribution from the Milky Way
at the position of IC 10 as a function of V magnitude and V$-$I color.
We transformed B$-$V in the model program to the Kron-Cousins V$-$I
using the assumption that the luminosity class of the stars is V or III 
(not I) as indicated by the model itself. 
The authors suggest 25\% uncertainties in predicting star counts, but
also warn of increased uncertainties for $b<10$\arcdeg. IC 10
is at a $b$ of $-$3.3\arcdeg, and thus, the model results are uncertain. 
The model predicted the number
of stars per unit area in different color bins for V magnitudes of 18 to 28.

To check the model,
we used the ground-based images of IC 10 provided by P.\ Massey to estimate
the contribution of stars in the Milky Way to the star counts of IC 10.
We measured the V-band brightnesses and V$-$R colors of the stars outside
IC 10 (beyond a radius of 5.6\arcmin\ centered on IC 10), 
determined incompleteness
corrections, 
constructed a corrected luminosity function, and converted V$-$R to V$-$I.
The ground-based image covers only V magnitudes to 22; fainter than 
that incompleteness
factors exceed 50\%. However,
in the brighter 3.5 magnitudes that we measured,
the results from the ground-based images and predictions from the Bahcall-Soneira
model agree very well.

Therefore, when we wished to examine the entire galaxy population
or the underlying galaxy population,
we used the Bahcall-Soneira model predictions to subtract the foreground
stellar population. However, when we wished to isolate the recent 
starburst, we used the {\it HST} images themselves to subtract the
underlying galaxy population as well as foreground stars. We did this by
identifying the region not included in the starburst, mostly WF2 and
WF3; counting stars;
correcting for incompleteness; scaling by the relative areas of the
chips; and subtracting the numbers in the underlying galaxy from 
those in the burst region. This assumes that the burst region has
undergone the same star formation history as the region outside of that
until the burst began.

\section{Stellar Initial Mass Function} 

The stellar mass function is the number of stars {\it counted} as a function
of the mass of the star. The stellar {\it initial} mass function, or IMF, is
the number of stars {\it formed} as a function of the mass of the star.
Obviously, it is the IMF that we want to determine
in order to probe the star formation process.
We assume that the
mass function is a power law, $f(m)=Am^{\gamma}$, in the notation of
Scalo (1986). Then $\xi(\log m) = (\ln 10) m f(m)$ and
$\Gamma = {{\partial \log \xi(\log m)}\over{\partial \log m}}\vert_m$,
where $\gamma=\Gamma-1$.
The slope of the IMF that we measure is $\Gamma$, 
and for a Salpeter (1955) IMF, 
$\Gamma$ is $-$1.3.

For a coeval population that is younger than the lifetimes of
the stars being considered,
the IMF is constructed by counting the number of stars in
mass bins and dividing by the difference between the logarithm of the 
mass that brackets the upper end of the bin and the logarithm of the
mass that brackets the lower end of the bin. The mass bins are chosen
to be approximately equal in this difference. One often then divides
by the area of the galaxy being surveyed although we will not 
do that here since it does not affect the power law index.
The power law index, refered to as the ``slope'' of the IMF, is 
then the slope $\Gamma$ of the best linear fit
to the logarithm of the number of stars per mass bin $\xi$ versus
the logarithm of the average stellar mass of the bin.
We take the uncertainty in $\xi$ as the root-N statistics
of the number of stars counted, and the uncertainty in $\Gamma$ as the
uncertainty in the linear fit to $\log \xi$ versus $\log m$.

If a stellar population is not coeval, one must consider
the star formation history of the region. In the case where star
formation has been constant, for stars with hydrogen-burning lifetimes
that are less than the age of the region, Scalo (1986) shows that 
$\xi(\log m) = {{\phi_{ms}(\log m) T_0}\over{\tau_{ms}(m) b(T_0)}}$,
where $\phi_{ms}(\log m)$ is the present-day mass function determined from the 
luminosity function
of main sequence stars, $T_0$ is the age since star formation began in the
region, $\tau_{ms}(m)$ is the hydrogen-burning lifetime of a star of mass $m$,
and $b(T_0)$ is the birthrate at $T_0$ relative to the average over the
region's lifetime.
This assumes that the IMF
has been independent of time. 
The ratio $T_0$/$\tau_{ms}(m)b(T_0)$ serves to scale the number of
stars counted today in order to correct for stars now dead,
and if we assume that $b(T_0)=1$, that is, a constant star formation rate,
the scaling factor is just
the ratio of the age of the region to the hydrogen-burning lifetime
of the stars.
Thus, since $T_0$ is a constant, the effect of assuming constant 
star formation is to divide the coeval mass function by the 
hydrogen-burning
lifetimes of the stars in each mass bin.

In what follows, we have used the isochrones and stellar evolutionary
tracks compiled by
Lejeune \& Schaerer (2001). 
The oxygen abundance in IC 10, determined from \HII\ regions,
is measured at log(O/H)$+12=8.26$ (Garnett 1990), which implies
a $Z$ of 0.005. 
We, therefore, use isochrones and stellar evolutionary tracks
for metallicities of $Z=0.004$ and $Z=0.008$.
The stellar hydrogen-burning lifetimes, time-weighted average M$_V$, and U, V,
and I isochrones
were determined or taken from
Lejeune and Schaerer's basic models with standard stellar mass-loss.

\subsection{Starburst Region}

\subsubsection{Color-magnitude Diagrams} \label{seccmd}

Because we wish to determine whether the stellar IMF or stellar lower mass
limits were unusual during the recent starburst, we have isolated the portion
of the galaxy in the {\it HST} field-of-view that appears to have taken part
in this starburst. We drew a boundary around this region by looking for 
the portion of the field-of-view that is higher in surface brightness than
its surroundings and where there is a higher density of bright, blue stars.
The region thus chosen is outlined in Figure \ref{figburst}.
The underlying galaxy was taken to be the regions not included in the
starburst.
A dark cloud in WF4 was excluded from consideration in either 
category and is outlined
with a box in Figure \ref{figburst}.
The starburst is located primarily in PC1 and WF4, with small pieces
in WF2 and WF3. The region contains an area equivalent to a square
425 pc on a side.

Color-magnitude diagrams of stars in the burst region and in the
underlying galaxy
are shown in Figures \ref{figcmdburst} and
\ref{figcmdback}.
Superposed are isochrones for $Z=0.004$ metallicity from Lejeune
\& Schaerer (2001). The zero age main sequence of the isochrones
falls along the blue edge of the blue plume, as expected.
In the CMD of the starburst region we see 
that the red extension of the isochrones to the red
supergiants falls of order 0.1---0.2 magnitude in (V$-$I)$_0$
short of many of the observed stars. 
We use $Z=0.004$ stellar
evolutionary tracks because this metallicity is close to the oxygen abundance
determined for IC 10 from nebular emission. 
However, the more metal-rich $Z=0.008$ isochrones extend further
to the red than $Z=0.004$ isochrones and so encompass the observed
red supergiants without changing significantly the blue main sequence.
In \S\ref{secclcolor} we will also find that $Z=0.008$ cluster
evolutionary tracks (derived from the same stellar evolutionary tracks)
better match the observed integrated colors of star clusters in
IC 10. 
Since the oxygen abundance in IC 10 implies $Z=0.005$, one would expect
the $Z=0.004$ tracks to be a little too metal-poor, and probably the
best fit would be tracks between $Z=0.004$ and $Z=0.008$.
Thus, we have calculated the stellar IMFs in IC 10
using both $Z=0.004$ and
$Z=0.008$ stellar evolutionary tracks. 
However, this mismatch of observed $Z$ and evolutionary tracks
is not unique to IC 10; the same issue
arose in studies of clusters in two other metal-poor irregulars---
NGC 1569 (Hunter \et\ 2000) and NGC 4449 (Gelatt, Hunter, \& Gallagher
2001) where higher metallicity cluster evolutionary tracks fit
observed cluster colors better.

\subsubsection{The Stellar IMFs} \label{secimf}

Although a starburst represents an unusually high star formation rate
over a short period of time, it is not clear whether this starburst
took place over a short enough period of time to be considered coeval
or whether it must be considered constant star formation over some
extended time interval. By coeval, we mean that the duration of the event
is less than the hydrogen-burning lifetimes of the most massive stars 
being considered, here 11 Myr for 18 M\solar\ or 13 Myr for 
the lifetime of the average mass
in the top mass bin.
The CMD suggests that star formation 
may have continued over a period of several tens of millions of years,
extending to as recent as 4 Myr ago, but the WR data may suggest
that star formation took place over a very short period of time. 
Because of the uncertainty of how the starburst proceeded, we have determined 
the IMF for the two limiting extremes: coeval and constant star formation.
These IMFs should bracket the true IMF if the star formation history
during the burst was more complex.

We have determined the IMF in 4 mass bins from 6.3 M\solar\ to
18 M\solar. The masses of stars more massive than 18 M\solar\ cannot
be determined from photometry alone (Massey 1998)
and so are not included here. 
Stars in the next lower mass bin, 4.8--6.3 M\solar, are not included
because incompleteness for this mass bin are 68\%--92\% in the
burst region.

As suggested by Scalo (1986), we only
include stars on the main sequence, which we take to be (V$-$I)$_0\leq0.24$.
Stars with masses 6.3--18 M\solar\ remain at (V$-$I)$_0<0$ while on
the main sequence, but we take a larger color swath here to allow for
uncertainties in the photometry of the colors and reddening corrections
and to include the entire blue plume.

We must determine the relationship between mass and M$_V$ for the limits
of our mass
bins. Since M$_V$
varies as a star evolves even while the star is in its hydrogen-burning phase,
we determined a time-weighted M$_V$, but the time-scale that we average
over depends on our assumptions.
Stars with masses 6.3--18 M\solar\ have hydrogen-burning lifetimes of 
11 to 62 Myr. 
For the case of coevality, we assumed that the star formation 
took place over a time period $\leq$13 Myr,
the lifetime of the average mass in the top mass bin.
Therefore, to convert M$_V$ to mass, we have used M$_V$ averaged
over 13 Myr.
This should then fairly reflect
the average M$_V$ that we would expect for a given mass under our assumptions.
The adjustment from zero-age M$_V$ to average M$_V$ is 0.6 
magnitude brighter for 18 M\solar\ and decreases with decreasing mass
(increasing lifetime).

For the case of constant star formation, we assumed that 
the age of the region $T_0$ is 40 Myr and
that star formation has taken place at a constant rate there
since then to the present. 
The choice of 40 Myr is motivated by the
CMDs in which we see a strong blue main sequence as well as
red supergiants on isochrones consistent with ages of up to 40 or 50 Myr.
Ages of clusters, discussed below, are also consistent with this.
In the constant star formation case we must multiply the mass function by
$T_0$/$\tau_{ms}(m)$ 
in order to account for those
stars that have died more than $\tau_{ms}(m)$ ago.
We determine the average M$_V$ over the hydrogen-burning lifetime
for each mass limit. 
The exception is the lower mass limit of the lowest mass bin; the
hydrogen-burning lifetime
of a 6.3 M\solar\ star is 62 Myr, which is longer than $T_0$, so we
average M$_V$ over $T_0$ rather than 62 Myr.

The IMFs for these two cases are shown in Figure \ref{figburstimf}
and given in Table \ref{tabimf}. 
The resulting $\Gamma$ are $-1.9\pm0.4$ under the assumption of
coevality and $-0.9\pm0.3$ under the assumption of constant star formation
for $Z=0.004$.
Using $Z=0.008$ stellar evolutionary tracks yields $\Gamma$ of
$-2.1\pm0.4$ and $-1.0\pm0.04$, respectively, nearly the same within
the uncertainties.
The real situation must lie between these two extremes. For comparison
a Salpeter (1955) slope is $-1.3$. Thus, in the case of coevality, the
IMF is steeper than Salpeter (fewer higher mass stars relative to lower
mass stars), and in the case of constant star formation, the IMF is
shallower than Salpeter. However, the uncertainties and likelihood of
the true IMF lying between these two extremes
do not rule out the IMF being the same as that measured by
Salpeter.
It seems likely, therefore, that the IMF of the intermediate mass
stars formed in the starburst is not very unusual compared to
what is found in most other populations.

\subsection{Underlying Galaxy}

We have also measured the IMF for stars in the region outside of the burst
under the assumption that the galaxy there has undergone constant star
formation for at least the past 100 Myr, the hydrogen-burning lifetime
of 4.8 M\solar, our lower mass limit. 
We included the mass bin 4.8--6.3 M\solar\ here, because in the underlying
galaxy the
incompleteness factors for this mass bin are lower than they
were in the starburst region, 57\%--67\%.
Because we are dealing with the constant star formation case,
we must multiply the stars counted by $T_0$/$\tau_{ms}(m)$ in order to
account for stars that have died longer than $\tau_{ms}(m)$ ago.
We take $T_0$ to be 10 Gyr although, because $T_0$ is a constant in
the IMF, the particular age chosen is irrelevant to determining the
slope of the IMF.
Again, we determine the 
time-weighted average M$_V$ of each mass bin limit over
the hydrogen-burning lifetime of each mass.
We considered the regions
in WF2 and WF3 not included in the starburst (see Figure \ref{figburst}).
The area included is equivalent to a square 375 pc on a side.
To remove foreground stars, we used the Bahcall-Soneira model as described
above. 

The IMF for the non-burst region is shown in Figure \ref{figoldimf}
and given in Table \ref{tabimf}. 
$\Gamma$ is $-2.6\pm0.3$ for 4.8--18 M\solar\ for $Z=0.004$.
Using $Z=0.008$ stellar evolutionary tracks yields a $\Gamma$ of $-2.3\pm0.3$.
This is considerably
steeper than the Salpeter (1955) function that is commonly measured
in stellar clusters and associations in a wide variety of galaxies.
The deviation is in the sense of fewer higher mass stars for a given
number of lower mass stars. However, slopes this steep have been measured
in a few situations:
The slope is similar,
within the uncertainties, to the slope measured for the starburst
region in IC 10 under the assumption of coevality.
It is also similar to the slopes measured by Massey \et\ (1995b)
for field populations of massive stars in the LMC, SMC, and Milky Way
analyzed in the same manner.

If the star formation history of the underlying galaxy has been episodic
over the past 100 Myr rather than constant, from the paucity of
bright blue stars and red supergiants, we would guess that the
last burst must have ended of order 20 Myr ago. This corresponds to
the main sequence lifetime of an 11 M\solar\ star. If we eliminate
bins for masses greater than 10.7 M\solar, the IMF becomes even
steeper although uncertainties are large.
Thus, the unusual character of the IMF remains.

\section{Lower Stellar Mass Limit}

We measure the IMF to 6.3 M\solar\ in the starburst region.
Thus, the lower stellar mass limit is $\leq$6.3 M\solar,
which corresponds roughly to a B3V star.
We detect stars down to 4.8 M\solar\ (B5V), but incompleteness is so high
in the 4.8--6.3 M\solar\ mass bin (68\%--92\%)
that detetermining numbers of stars is too uncertain to
say whether they are there in normal proportions.
The constraint of $\leq6.3$ M\solar\
is less than the 10--20 M\solar that is suggested for
some systems from
integrated measurements (for example, Augarde \& Lequeux 1985, Olofsson 1989)
and from theory (Silk 1986),
but lower than others (for example, Rieke \et\ 1993, Wright \et\ 1988).

\section{Wolf-Rayet Stars and the Unusually high WC/WN ratio} \label{secwr}

Massey \& Armandroff (1995) surveyed IC 10 for 
WR stars and 
found an unusually large ratio of WC-type stars to WN-type stars.
The WR stars represent an evolutionary phase in the life
of massive stars where strong mass loss has laid bare nucleosynthesis
products: helium and nitrogen in the case of WN stars and carbon and
oxygen in the
case of WC stars.
Recently, Royer \et\ (2001) have also surveyed IC 10 for WR stars and
detected 13 new WR candidates. If confirmed with spectroscopy,
8 of the new WR stars would be of the WC type, and so the
abnormally high WC/WN ratio will likely remain.

The WFPC2 field of view of IC 10 includes 6 of the 10 WC stars and 3 of the 5
WN stars, as well as one of the WN candidates, identified by
Massey, Armandroff, \& Conti (1992). The final list of WR stars and
their coordinates is given by Massey \& Johnson (1998).
The field also contains 6 of the new WR candidates of Royer \et\ (2001).
The WR stars and candidates are marked on Figure \ref{figburst},
although the Royer \et\ candidates numbered 11 and 12 do not
correspond to any star in our image.
All but one WR star, found in WF2, are located
within the region taken as the burst region.
Only one is located in close proximity to an
obvious grouping of stars (cluster 2--1).

The WC/WN ratio of 2 found in IC 10
is 20 times too high for the galaxy's metallicity (Massey 1999).
Massey \& Armandroff (1995) had suggested that the starburst in IC 10 may 
have an IMF that is skewed towards the highest
mass stars although the IMF would have to be extraordinarily peculiar
to alone explain this ratio. Recently the work by
Massey, Waterhouse, \& DeGioia-Eastwood (2000) has shown that the same mass range contributes 
stars to the WC and to the WN type.  However, if the time spent
in the WC phase increases with mass, as expected, the unusual WC/WN ratio
could still be due to the stellar IMF.
However, to be explained in this fashion, there would have to be
more of, say, 100 M\solar\ stars than 40 M\solar\ stars.
That is, the IMF slope, if the IMF was still a power law,
would have to be positive, not negative.
Thus, the IMF that we have found for the intermediate mass stars, if it extends
to high mass stars, is not 
extraordinary enough to explain the observed WC/WN ratio.

Massey \& Johnson (1998) suggest that one explanation for the unusually
high WC/WN ratio in IC 10 could be that the mixed-age assumption 
in the empirical relationship is violated in IC 10. The empirical
relationship shows WC/WN versus O/H for galaxy-wide counts, and the
assumption is that star formation has been constant on a galaxy-wide scale
in all these galaxies.
On the other hand, if a galaxy population is highly coeval, as it might
be in a starburst, the WC/WN ratio reflects the time since
the burst occurred. 

This has been quantified nicely by
Schaerer \& Vacca (1998) who show the evolution of WC and WN fractions
with time for an instantaneous burst of star formation.
For $Z=0.004$, close to the $Z=0.005$ expected for IC 10,
the WC/WN ratio is greater than 1, and in fact
about 2, 3--4 Myr after the burst. However, this would require that
the burst in IC 10 be only 3--4 Myr old everywhere in the galaxy since
the WR stars are found over a large region. This
seems unlikely given the large number of red supergiants that are seen
in the CMD (Figure \ref{figcmdburst}); an age or burst duration of 
several tens of millions of years seems more likely.

However, Figure 9 of Schaerer, Contini, \& Kunth (1999) is also suggestive.
They have computed the expected WC/WN ratio for an instantaneous burst,
bursts of several Myr in duration, and constant star formation, 
including several metallicities and both standard and high stellar mass
loss. For the standard mass loss and $Z=0.004$ the highest WC/WN ratio
attained is only 0.3 and then only for a very short period of time.
The longer the burst duration, the lower the peak ratio, and the
WC/WN ratio for constant star formation is 0.04. Clearly these values
fall far short of what is observed in IC 10. 

However, the stellar evolutionary models with high mass-loss can produce
much higher WC/WN ratios.
Again, the instantaneous burst and constant star formation models
only produce ratios of order 2 for ages 3--4 Myr.
The constant star formation model settles down to WC/WN of order 0.5
after 4 Myr.
However, the uncertainty in the measured WC/WN ratio in IC 10 is
such as to allow a value of 1 (Massey 1999), and Massey \& Johnson (1998),
as well as Royer \et\ (2001),
suggest that incompleteness in the WN count is a possibility.
If the true WC/WN ratio in IC 10 were closer to 1, the
high stellar mass loss models with constant star formation 
might nearly predict what
is observed. However, this is a circular argument
because mass-loss rates are related to the metallicity
of the stars. So, we
would require that the stars in IC 10 have unusually high mass-loss
rates for their metallicity compared to the 
other galaxies that define the metallicity--WC/WN relationship.
There is no reason to expect this to be the case.

More likely, the WC/WN ratio is high because of a narrow range in
ages as shown by Schaerer \& Vacca (1998). The ratio
could be reproduced 
if the bulk of the starburst occurred some tens of
millions of years ago, there was a gap in star formation until recently,
and then scattered pockets of star formation 3--4 Myr ago produced the 
WR stars that we see now. 
This is also suggested by Royer \et\ (2001).
In this scenario the numerous
red supergiants
were produced in the older burst and are decoupled from the production 
of the WR stars that we see today. This does require, however, 
a highly synchronized second burst ($\Delta\tau\leq$1 Myr) in small pockets
scattered over a large region of the galaxy.
In terms of the IMFs computed in \S\ref{secimf}, these pockets
are too small to contribute significantly to the star counts of the
starburst region as a whole.

\section{Star Clusters} \label{seccl}

\subsection{Identification}

We have examined the WFPC2 images for star clusters or distinct
associations. In particular we were looking for objects that might
be comparable to super star clusters, such as R136 in the LMC, or
the smaller but more common populous clusters, also found in the LMC.
R136, for example, has a half-light radius \rhalf\ of 
1.7 pc (Hunter \et\ 1995), which would be 0.4\arcsec\ at the distance
of IC 10. Thus, a cluster like R136 would not be distinguishable
from a star in ground-based images, but in {\it HST} images
the \rhalf\ of R136 would be 8 pixels in PC1 and 4 pixels in the WF CCDs,
easily distinguishable from a stellar profile.
Therefore, to search for compact clusters, we looked for anything that 
was resolved relative to a stellar profile.
We also looked for less compact clusters or associations.
Because of the starburst, however, much of the galaxy looks like 
an OB association. We looked only for those
associations that appeared as 
distinct clumpings of stars.
In what follows we will frequently refer to both clusters and associations that
we identified as ``clusters.''
We initiated our search on the F555W images, but then compared
that image to F336W and F814W. However, we did not restrict ourselves to 
blue clusters, and the reddest clusters do not show up on the F336W
image.

We identified 3 bright, blue star clusters and 1 faint, red cluster
in the PC1 image; two clusters in WF2; and 7 clusters in WF4.
No clusters were identified in WF3.
The clusters are outlined in Figure \ref{figburst} and their
properties are given in Table \ref{tabclusters}.
We identify the clusters with a number with two parts: the leading
part is the number of the CCD that the cluster is found on
and the following part is a running number to uniquely identify 
the cluster on that chip.

The clusters have half-light radii \rhalf\ of 1.5--6.6 pc and
M$_V$ of $-$6.6 to $-$10.
Clusters 4--7 and 4--8 are identifiable
as resolved objects, but they are saturated in the 
F555W and F814W images and so have no photometry or \rhalf\
listed in Table \ref{tabclusters} and cannot be discussed further.

\subsection{Integrated Colors and Magnitudes} \label{secclcolor}

The integrated colors and magnitudes of the clusters are given
in Table \ref{tabclusters}. The clusters are also plotted
in color-color diagrams and CMDs in Figures \ref{figclccd}
and \ref{figclcmd}. In the Figures we also plot
cluster evolutionary models for $Z$ of 0.008 and 0.004 and a Salpeter IMF
(Leitherer \et\ 1999). Ages along the evolutionary tracks of
1--9 Myr in steps of 1 Myr are marked with
x's and ages of 10, 20, and 30 Myr are marked with open circles. The
cluster evolutionary tracks end at 1 Gyr.
Although a $Z$ of 0.004 is closer to the expected metallicity of
IC 10, the $Z=0.008$ track appears to follow the observed
colors better in Figure \ref{figclccd}. 
In particular cluster 1--1 lies close to the evolutionary cluster
track of $Z=0.008$ and far from that of $Z=0.004$.
Better agreement of integrated cluster colors with a higher
metallicity evolutionary track was also the case
for clusters in NGC 1569 (Hunter \et\ 2000) and in NGC 4449 (Gelatt,
Hunter, \& Gallagher 2001).

Of the clusters with measureable F336W, inferred ages span a
range from 3--4 Myr to about 30 Myr, allowing for the uncertainties, 
and so they were probably produced
in the starburst. Cluster 2--1 has an uncertain
(U$-$V)$_0$ that could formally allow an age of $\sim$7 Myr to
200 Myr. However, it contains a WR star (Figure \ref{figburst})
and so must be young. 
In addition to the uncertainties in the photometry, however,
there is also considerable uncertainty
in determining ages of small clusters from integrated photometry
because of the statistical effects of a few stars entering and leaving particular
evolutionary
stages. However, for those clusters large enough spatially to be resolved 
into individual stars, the integrated colors are consistent with 
the ages derived from CMDs. 
CMDs of two resolved clusters are presented in the next section.

Clusters 1--4 and 2--2 are red and do not appear
in Figure \ref{figclccd} because F336W was not detected. 
However,
they do appear in the bottom panel of Figure \ref{figclcmd}.
Given that F336W was not detected, we estimate that (U$-$V)$_0\geq0.2$.
From this and the observed (V$-$I)$_0$, we conclude that
the ages of 1--4 and 2--2 are likely
$\geq$350 Myr and could be
as high as 1 Gyr since the clusters sit near the end of the
cluster evolutionary track.

The cluster evolutionary tracks in Figure \ref{figclcmd} are
for a cluster mass of 3.3$\times10^5$ M\solar.
For less massive clusters the tracks
in Figure \ref{figclcmd} would slide vertically to fainter M$_V$.
We can see that, if the ages are correct, most of the clusters in IC 10 
have masses of a few thousand M\solar\ or less and are, in fact, very small
clusters or associations. 
The young clusters 4--1 and 1--2 and the old clusters
1--4 and 2--2 could have masses as high $10^4$ M\solar. This mass
would still be small compared to that of a globular cluster or 
super star cluster, but perhaps not too much smaller than smaller
populous clusters in the LMC. For example, the populous cluster
NGC 1818 in the LMC has a total mass determined from star counts at 20 Myr 
and then integrated
from 0.1 to 100 M\solar\ of $3\times10^4$
M\solar\ (Hunter \et\ 1997).

The spatial distribution of star clusters can be seen in Figure \ref{figburst}.
The youngest clusters (4--1, 4--2, 4--5, 2--1) are primarily located
in the upper left corner of WF4, in the region with the most nebular
emission. The exception is cluster 2--1 that is potentially only a few Myr 
older and
is located along the right edge of WF2. The rest of the young clusters
are arrayed around 4--1, 4--2, and 4--5, to the lower right and in
the PC1, and have ages of order 15--30 Myr.
Although invoking propagating star formation is tempting,
we cannot in all honesty say that this arrangement necessarily implies 
causality between regions.

\subsection{Resolved Clusters} \label{secrescl}

The CMDs of clusters 1--1 and 4--1 
are shown in Figures \ref{figcl11cmd} and \ref{figcl41cmd}.
Comparison with isochrones suggests that cluster 1--1 could be
15--20 Myr old. The integrated colors suggest an age of 
10--14 Myr, so the two ages agree at an age of about 15 Myr. 
Cluster 4--1  appears to be
as young as 4 Myr, and the integrated colors also indicate
an age of 3--4 Myr.

For cluster 4--1 we have determined the
richness (number of stars)
and spatial concentration (number of stars formed per unit area) of stars
with M$_V\leq-4$, in other words all massive stars.
Star counts determined within the cluster boundary were corrected for incompleteness and 
contamination by foreground stars
(which is $<$0.2 star in each magnitude bin under consideration).
In Figure \ref{figdensity}
we compare this cluster with other clusters of
similar age. We see that cluster 4--1 is comparable to an OB association.
It is not as concentrated or as rich as a super star cluster, as exemplified
by R136 in the LMC, by several orders of magnitude.
In fact, no cluster that we identified in the field of view of
the {\it HST} images is comparable to a super star cluster.

\section{H$\alpha$ Shells} \label{secha}

Figure \ref{fighmrmos} is a mosaic of the \ha\ CCD images with stellar
continuum subtracted to leave just nebular emission. 
One can see that there is a considerable amount of \ha\ emission
with a complex pattern of filaments
in WF4. 
This emission indicates that it is here that the most recent star
formation has taken place.
The \HII\ regions are numbered and catalogued by Hodge \& Lee (1990).

A large shell dominates the \ha\ image and is seen 
in the upper left corner of WF4. 
This is shown more clearly in Figure \ref{fighmr} where the \ha\ image
is superposed on the F555W image for the WF4 CCD only.
The large shell is number 111 in Hodge \& Lee's (1990) catalog.
Cluster 4--1, outlined in the Figure, is located along
the upper edge of this shell, where the \ha\ emission is strongest.
Cluster 4--2 is located towards the center of the shell.
The CMDs and integrated colors of these clusters indicate ages
of a few Myr for both.
The \ha\ shell is 9.7\arcsec, or 45 pc, in diameter. 
If the ambient gas density $n_0$ is 1 cm$^{-3}$, a single
massive star could easily blow a hole this size in a few
Myr (Weaver \et\ 1977, McCray \& Kafatos 1987), so there is
no problem producing this shell with stars in clusters 4--2 or
4--1.

To the lower right of this shell in Figure \ref{fighmr}
are bright arcs of \ha\ emission.
Cluster 4--4 sits near the center of curvature of several of these
arcs (\HII\ number 106 in Hodge \& Lee 1990) which together look something
like a spiral galaxy.
The distance from the cluster to the outer arcs is 5.4\arcsec$=$25 pc.
Cluster 4--4 has integrated colors that indicate an age of 20--30 Myr,
and this may be an older shell in the process of breaking up.
Another tiny shell is found to the right of cluster 4--3
(\HII\ number 98 in Hodge \& Lee 1990).
That shell is centered on
a star and has a radius of 1.68\arcsec$=$8 pc.

Thus, we find that in \ha\ the shells are quite modest. This implies
that in this part of the starburst region there are no super shells
nor is blowout of gas from the galaxy likely (but see Hunter \et\ 2001
for other parts of the galaxy).

\section{The Mode of Star Formation in the Starburst}

From the discussion in the previous sections we see that, in terms
of star clusters,
the starburst in IC 10 has not produced
super star clusters but rather normal OB associations and small
compact clusters. Only two of the young clusters may be comparable to 
smaller populous clusters such as are observed in the LMC.
By contrast, the starburst in NGC 1569 has produced two super star clusters
(see, for example, O'Connell, Gallagher, \& Hunter 1994)
and that in NGC 4449 has produced several as well (Gelatt \et\ 2001).
Yet, IC 10 is not unique in being a region of intense star formation and
not producing super star clusters:
the Blue Compact Dwarf galaxies
IZw18 (Hunter \& Thronson 1995)  and VIIZw403 (Lynds \et\ 1998)
also have produced only scaled-up OB associations.
Clearly, not all starburst events produce super star clusters
even if 
giant \HII\ regions are present
(see also Kennicutt \& Chu 1988), and the star formation
process in IC 10 has been similar to that in IZw18 and VIIZw403
in terms of cluster production.

In the case of IC 10, the primary mode of star formation during
the starburst appears to have been that of 
an OB association.  Here we use the phrase ``mode of star formation''
to mean the primary units of star formation.
The region identified as the burst region
in Figure \ref{figburst} is similar to an OB association in
the distribution of stars. This is seen in Figure \ref{figburstden}.
This plot is similar to Figure \ref{figdensity} in which we plotted
the richness of stars in cluster 4--1 against the 
spatial concentration of those
stars. In Figure \ref{figburstden} instead
we plot the richness versus the spatial concentration for stars in the
burst region as a whole. In addition, we count stars with masses
6.5--15 M\solar\ rather than stars with M$_V<-4$.  
The advantage of counting stars in this mass range rather than to
a brightness limit is that one can compare regions that are not the same
age. The only requirement is that the regions be younger than the
hydrogen-burning lifetime of the upper mass limit, here 
a 15 M\solar\ star---about 13 Myr, or that we can reasonably
correct for stars that have evolved off the main sequence, as
for NGC 1818, which is shown for comparison in the figure.
For IC 10 we have counted stars 6.3---18 M\solar, corrected for
incompleteness and foreground stars, and scaled to 6.5---15
M\solar\ using the observed mass function.
The IC 10
burst region is compared to other star-forming regions in the figure.

From Figure \ref{figburstden}
we see that the spatial concentration of stars in the IC 10 burst
region is that of a typical OB association seen in the Milky Way and
LMC, but the numbers of stars
are up by a factor of 40--400 compared to OB associations.
These stars are located over a region equivalent to a
square with a side of 425 pc or a circle with radius 240 pc. 
In fact, the mode of star formation
in IC 10 is similar to that in the bulk of the stars in 
Constellation III in the LMC (labelled ``ConIII-field'' in the Figure).
Constellation III is a large region that formed stars 12--16 Myr
ago (Dolphin \& Hunter 1998). 
The stars have now carved a hole out of the neutral gas that
is $\sim$1 kpc in diameter (Dopita, Matthewson, \& Ford 1985).
In Constellation III there are several compact clusters, but the
bulk of the stars that formed in what was certainly a major star-forming event
formed in the more diffuse distribution of an OB association.
This appears to be the situation in the IC 10 starburst as well.
The numbers of stars formed is also similar between the region
surveyed by {\it HST} in IC 10 and the region surveyed from
ground-based observations in Constellation III by Dolphin and Hunter.
Both Constellation III and the starburst
region surveyed in IC 10 cover roughly the same fractional area
of their respective galaxies. Besides the similarity with Constellation
III, IC 10's mode of star formation---that of OB associations---is
like those in the two Blue Compact Dwarfs IZw18 and VIIZw403
(Hunter 1999).

\section{Summary}

We have presented {\it HST} U, V, I, and \ha\ images of the peculiar Local Group 
irregular galaxy IC 10. The images are used to determine the nature
of the stellar products of the recent starburst in this galaxy.  
From these products
we have probed the star formation process of the starburst itself.
The starburst is explored through a region of the galaxy equivalent
to a square with a side of 425 pc. The region outside of this is taken to
represent the underlying galaxy population.

We assumed a single reddening of E(B$-$V)$=0.77$ and found that this
gave CMDs that compare well to isochrones.
From the TRGB, we deduced a true distance modulus
of 24.95$\pm0.2$, thus reproducing the reddening and distance combination
of Massey \& Armandroff (1995).

We identified 13 stellar associations and clusters in the {\it HST}
images. Two of these are red and presumably old ($\geq$350 Myr). Integrated colors
and CMDs suggest ages of 4--30 Myr for the rest and these were presumably
formed in the starburst. The youngest associations and clusters
are located in the strongest part of the nebulosity; the older clusters
are located around the younger ones and within the starburst region.

The answers to the questions that we posed in the Introduction are as follows:

First, what is the stellar IMF of the intermediate mass (6.3--18 M\solar)
stars produced
in the starburst?
We determined the IMF for two limiting cases. Under the assumption
that the starburst was highly coeval and took place no more than 13 Myr
ago, the slope of the IMF is $-1.9\pm0.4$.
Under the assumption of constant star formation over the past 40 Myr,
the slope is $-0.9\pm0.3$. The true case is expected to lie somewhere
between these two limits. 
These are calculated using 
$Z=0.004$ stellar evolutionary tracks, which is close to
the observed
metallicity of IC 10. For $Z=0.008$
tracks, which fit the location of the red supergiants better,
$\Gamma$ is $-2.1\pm0.4$ and $-1.0\pm0.04$, respectively, for 
coeval and constant star formation.
Thus, most likely, the IMF of the intermediate mass stars is not
very unusual.
We also measured the IMF for the 
underlying galaxy population under the assumption of constant
star formation for more than 100 Myr. We found a slope of $-2.6\pm0.3$
for 4.8--18 M\solar\ assuming $Z=0.004$ ($-2.3\pm0.3$ for $Z=0.008$)
which is unusually steep compared to values measured
in most circumstances in other galaxies.

Second, what is the lower stellar
mass limit of stars produced in the recent starburst?
The lower stellar mass limit is $\leq$6.3 M\solar\ because
we measured stars in relatively normal proportions to this limit.
This is an upper bound to the lower stellar mass limit; we are constrained
in pushing this bound lower by the signal-to-noise of the data.
This constraint is less than some predictions of what lower stellar mass limits
might be in starbursts, but higher than others.

Third, what kinds of star clusters have been produced in the starburst?
We found a few distinct OB associations and small compact clusters. 
There are no super star clusters in the {\it HST}
field of view such as have formed in the starburst in NGC 1569.
At most, a few of the clusters may be comparable
to small populous clusters in the LMC.

Fourth, how does the nebular emission relate to the stellar products?
There is a lot of filamentary structure in the \ha\ image, but only
two modest-sized shells ($\sim$50 pc diameter). The shells are centered
on star clusters/associations that could easily have produced
them in the past few Myr. There is not evidence in this part of
the starburst for super shells or blowout activity.

Fifth, what has been the mode of star formation in the starburst?
By mode, we mean richness and concentration of the stars that
have formed.
The dominant mode of star formation has been that of a scaled-up
OB association. That is, the spatial concentration is like that
of an OB association, but altogether several orders of magnitude more
stars have been formed in IC 10 than one finds in typical OB associations.
This modest mode of star formation, 
with a few compact clusters sprinkled in, is similar
to the star formation that took place in Constellation III in
the LMC, as well as in the Blue Compact Dwarfs IZw18 and VIIZw403.

We also compare the high WC/WN ratio observed by Massey \& Armandroff
(1995) in IC 10 to evolutionary models of Schaerer \& Vacca (1998)
and Schaerer \et\ (1999). 
We suggest that 
the high ratio can be reproduced if there were small,
well-synchronized ($\Delta\tau\leq1$ Myr), 
but widely scattered, pockets of secondary star formation
3--4 Myr ago.

\acknowledgments

We are grateful for comments and suggestions
from P.\ Massey.
Support for this work was provided by NASA through
grant number GO-06406.01-A from the Space Telescope Science
Institute, which is operated by the Association of Universities
for Research in Astronomy, Inc., under NASA contract NAS5-26555.

\clearpage

\begin{table}
\dummytable\label{tabimf}
\end{table}

\begin{table}
\dummytable\label{tabclusters}
\end{table}

\clearpage

\figcaption{The field of view of the {\it HST} WFPC2 images of IC 10 is
shown superposed on a ground-based V-band image of the galaxy kindly
supplied by P.\ Massey. The field of view of the ground-based image
is 14.3\protect\arcmin.
\label{figfov}}

\figcaption{I-band color-magnitude diagram of stars in the non-burst portions 
of WF2 and WF3 taken to 
define the underlying stellar population. The magnitude and color have
been corrected for reddening, as discussed in the text. Stars to the
right of the dashed line, which include the RGB and AGB, were used in determining 
the I-band luminosity function. The solid, horizontal line is the apparent
magnitude that we determined for the TRGB.
\label{figicmd}}

\figcaption{I-band luminosity function, where the I magnitude has been corrected
for reddening. Stars were binned in 0.2 magnitude bins. The solid-lined histogram
is the counts corrected for incompleteness and foreground stars.
The other histogram is the counts before any corrections were applied.
The short vertical line at I$_0$$=$20.8 marks what we took to be the
TRGB.
\label{figlumi}}

\figcaption{Representative incompleteness fractions for star counts in
M$_V$ bins used
in \S 4.1, \S 4.2, and \S 7.3. These fractions 
are for main sequence stars, taken to
be stars with (V$-$I)$_0<0.24$.
\label{figincom}}

\figcaption{Region assumed to have engaged in the recent starburst is
outlined on a mosaic of the F555W {\it HST} images. The region consists of
the PC1, most of WF4, and portions along the right edge of WF2 and WF3.
The CCDs are identified in their corners.
The rest of the field of view was used for subtracting the underlying 
stellar population from the burst.
The large dust cloud in WF4, outlined with a box, was not used in either
category.
Star clusters and associations are circled and labeled. The size
of the circle is what is taken to be the maximum extent of the cluster.
Unlabeled circles indicate WR stars identified by 
Massey, Armandroff, \& Conti (1992).
The double circles are positions of WR candidates from Massey \& Johnson (1998)
and Royer et al.\ (2001) although two of the Roye et al.\ candidates
(one in PC1 and one in WF4) do not correspond to obvious stars in our 
images.
\label{figburst}}

\figcaption{Color-magnitude diagram of stars in the starburst region outlined
in Figure \protect\ref{figburst}. The uncertainties in the photometry are
shown vertically along the left edge of each panel for selected M$_V$ except where 
the uncertainties are smaller than the symbols.
Isochrones for $Z=0.004$ from 
Lejeune \& Schaerer (2001) are shown for selected ages which are 
identified in units of Myr.
The broadening in (U$-$V)$_0$ at faint M$_V$ in the right panel is
assumed to be partially due to red leak in the F336W filter that
was not properly accounted for.
\label{figcmdburst}}

\figcaption{Color-magnitude diagram of stars in the underlying galaxy. The
region is outlined
in Figure \protect\ref{figburst}. The uncertainties in the photometry are
shown vertically along the left edge of each panel for selected M$_V$ except where
the uncertainties are smaller than the symbols.
Isochrones with $Z=0.004$ from 
Lejeune \& Schaerer (2001) are shown for selected ages which are identified in units of Myr.
\label{figcmdback}}

\figcaption{Stellar IMF shown for the starburst region minus the underlying
galaxy. 
The IMF is determined for two assumptions about the star formation history
of the burst: coevality and constant star
formation. For coevality we require the star formation activity
to have taken place over a period of time shorter than the hydrogen-burning
lifetime of the most massive
mass bin being considered, 13 Myr.
For constant star formation, we assume that stars formed continuously over
the past 40 Myr.
\label{figburstimf}}

\figcaption{Stellar IMF shown for the non-starburst region. 
Foreground stars have been removed using the Bahcall-Soneira model as
described in the text.
We assume constant star formation over at least the past 100 Myr, the
hydrogen-burning lifetime of the lowest mass bin,
and an age of the galaxy of 10 Gyr.
\label{figoldimf}}

\figcaption{Integrated photometry of star clusters shown in a U, V, and
I color-color diagram.
The solid curve in the upper panel is an evolutionary track for a cluster with
instantaneous star formation, a metallicity of $Z=0.004$, and
a Salpeter (1955) stellar IMF with an upper
limit of 100 M\protect\solar\ (Leitherer \et\ 1999). The
solid curve in the lower panel
is the same models for a metallicity of $Z=0.008$.
Ages 1--9 Myr in steps of 1 Myr are marked with x's along these lines;
ages 10, 20, and 30 Myr are marked with open circles.
The cluster evolutionary tracks end at 1 Gyr.
The arrow in the lower left corner of the upper panel is a reddening line for
a change of 0.2 in E(B$-$V)$_t$. It represents the average of an O6 and
a K5 type spectrum with A$_V$/E(B$-$V)$=$3.1 and a Cardelli \et\
(1989) reddening curve.
\label{figclccd}}

\figcaption{Integrated photometry of star clusters shown in U, V, and
I color-magnitude diagrams.
The solid curve is an evolutionary track for a cluster with
instantaneous star formation, a metallicity of $Z=0.004$, and
a Salpeter (1955) stellar IMF with an upper
limit of 100 M\protect\solar\ (Leitherer \et\ 1999); the dashed
line is the same model for a metallicity of $Z=0.008$.
Ages 1--9 Myr in steps of 1 Myr are marked with x's along these lines;
ages 10, 20, and 30 Myr are marked with open circles.
The evolutionary tracks end at 1 Gyr.
The evolutionary tracks are those for a mass of 3.3$\times10^5$ M\protect\solar;
for other masses the lines would slide vertically
in the diagrams.
\label{figclcmd}}

\figcaption{Color-magnitude diagram of resolved stars within a radius
of 1.96\protect\arcsec\ ($=$9 pc) from the center of cluster 1--1.
The overplotted lines are $Z=0.004$ isochrones from Lejeune \& Schaerer (2001);
the numbers that label the lines
indicate ages in Myr.
\label{figcl11cmd}}

\figcaption{Color-magnitude diagram of resolved stars within a radius
of 3.09\protect\arcsec\ ($=$14 pc) from the center of cluster 4--1.
The overplotted lines are isochrones from Lejeune \& Schaerer (2001);
the numbers that label the lines
indicate ages in Myr.
\label{figcl41cmd}}

\figcaption{Number of stars with M$_V$ brighter than $-$4 
versus the spatial concentration of those stars in IC 10's
cluster 4--1 and other young regions.
Except for NGC 206, a region in M31 with
an age of roughly 6 Myr (Hunter \et\ 1996b), 
most of the regions listed are similar
in age to IC 10's cluster 4--1. Thus, we are comparing the
number of massive stars formed in star-forming events of similar
age. ``MW,LMC OB'' refers to OB associations in the Milky Way
and LMC (Massey \et\ 1995a,b); 
``R136'' is a super star cluster in the LMC (Hunter \et\ 1995).
Two regions
are shown in each of the Blue Compact Dwarfs IZw18 (Hunter \& Thronson 1995)
and VIIZw403 (Lynds \et\ 1998).
Data 
for NGC 604, a giant \protect\HII\ region in
M33, are taken from Hunter \et\ (1996a).
\label{figdensity}}

\figcaption{Mosaic of the 4 \protect\ha\ WFPC2 images.
Stellar continuum, formed from F555W and F814W, has been subtracted
to leave just nebular emission.
\label{fighmrmos}}

\figcaption{WF4 image. F555W is shown in black with 
F656N (\protect\ha) subtracted in order to superpose the \protect\ha\ as white. 
The star clusters
are outlined and numbered.
The positions of Br$\gamma$ sources numbered 3---6 of Borissova et al.\
(2000) are circled and labelled ``BR'' followed by the number of the
source in Borissova et al's Table 1.
\label{fighmr}}

\figcaption{Number of stars with masses 6.5---15 M\solar\ 
in the starburst region plotted
against the spatial concentration of those stars. The burst region
in IC 10 (``IC 10:burst'') is compared to other young regions.
``MW,LMC OB'' refers to OB associations in the Milky Way
and LMC (Massey \et\ 1995a,b), and the horizontal dashed line
extends the spatial concentration of typical OB associations
across the figure for comparison.
Data for R136, a super star cluster in the LMC, is taken from (Hunter \et\ 1995);
for NGC 206, an association in M31, from 
Hunter \et\ (1996b); for NGC 604, a giant \protect\HII\ region in
M33, from Hunter \et\ (1996a); and for NGC 1818, a populous cluster
in the LMC, from Hunter \et\ (1997).
``ConIII-cl'' and ``ConIII-field'' are clusters and field stars
in Constellation III in the LMC (Dolphin \& Hunter 1998).
\label{figburstden}}

\end{document}